\documentclass[12pt,preprint]{aastex}
\usepackage{natbib}
\shorttitle{Stellar Population of the Irregular Galaxy IC~10}
\shortauthors{{N. A. Tikhonov and O. A. Galazutdinova}}
\begin{document}
\makeatletter
\fontsize{12}{12} \selectfont

\title{Stellar Population of the Irregular Galaxy IC~10\footnote{Based on observations made with the NASA/ESA Hubble Space Telescope, obtained from the Data Archive at the Space Telescope Science Institute, 
which is operated by the Association of Universities for Research in Astronomy, Inc., under NASA contract NAS5-26555. These observations are associated with proposals 
6404, 7912, 9633, 9676, 9683 and 10242}}
\author{{N. A. Tikhonov and O. A. Galazutdinova}}
\affil{Special Astrophysical Observatory, Russian Academy of Sciences,N.Arkhyz, KChR, 369167, Russia}

\date{Received October 10, 2008}

\begin{abstract}
Based on our observations with the 6-m BTA telescope at the Special Astrophysical Observatory, 
the Russian Academy of Sciences, and archival Hubble Space Telescope images, we have
performed stellar photometry for several regions of the irregular galaxy IC~10, a member of the Local Group.
Distance moduli with a median value of $(m-M) = 24.47$, $D =780\pm40$ kpc, hav 
e been obtained by the
TRGB method for several regions of IC~10. We have revealed 57 star clusters with various masses and ages
within the fields used. Comparison of the Hertzsprung - Russell diagrams for star clusters in IC~10 with
theoretical isochrones has shown that this galaxy has an enhanced metallicity, which probably explains the
high ratio of the numbers of carbon and nitrogen Wolf-Rayet stars (WC/WN). The size of the galaxy's
thick disk along its minor axis is 10.5 and a more extended halo is observed outside this disk.

\end{abstract}

\section*{Introduction}
\indent \indent The irregular galaxy IC~10 is a member of the Local Group and is at a relatively close distance. However, 
it still remains a mysterious object in both morphology and physical parameters. Since the galaxy is
located in the Milky Way zone ($b = -3.\degr3$ ), its light undergoes strong extinction in gas-dust clouds of
our Galaxy. This creates difficulties in studying IC~10, because some of its characteristics are known to
within the extinction.

IC~10 ranks fitfth in luminosity in the Local Group, being less luminous than large spiral galaxies (M31,
MW, M33) and a bright irregular galaxy ---  the Large Magellanic Cloud. The integrated spectrophotometry
of IC~10 by \citep{hun85} revealed intense star formation processes in the galaxy. This is
also confermed by the images of IC~10 showing that the central galactic regions contain a large number of
bright young stars (\citep{vau65,kar93}. The existence of
bright and extended H II regions in the galaxy is also indicative of star formation activity (Hodge and
Lee 1990).

One of the peculiarities of IC~10 is a high space density of young Wolf-Rayet stars, exceeding that
in ordinary irregular galaxies (\citep{mas92,mas95,roy01,mas02,crow03}). In addition, an unusually large ratio of the numbers
of carbon and nitrogen (WC/WN) Wolf-Rayet stars is observed, which is in conflict with the low ($Z =
0.3Z_{\sun}$) metallicity of the interstellar medium in IC~10 found by \citep{leq79} and \citep{gar90}.

The presence of star formation over large areas of the galaxy, the high surface brightness, and the bluish
color (after correction for the galactic reddening) gave \citep{rich01} reason to believe IC~10 to be a
blue compact galaxy. This classification emphasizes that IC~10 is at a short stage of a starburst affecting
significant volumes of the galaxy.

Observations of the neutral hydrogen ($H~I$) in IC~10 were first performed by \citep{rob62}, who discovered
 a hydrogen cloud around the galaxy. The structure of this cloud with an angular resolution of $2\arcmin$
and its sizes were determined by \citep{sho74}, who found the cloud to be approximately elliptical
in shape and to rotate slowly. \citep{huch79} and \citep{coh79} ascertained that the sizes of the
hydrogen cloud reach one degree, which is hardly comparable to the galaxy's apparent sizes of 
$5.\arcmin9\times6.\arcmin8$\footnote{NASA/IPAC Extragalactic Database}. 
A detailed map of the H I distribution in IC~10 with an angular resolution of 5 was obtained by
\citep{wilc98}, who found that the structurally complex hydrogen cloud of IC~10 has shells
and holes produced by stellar winds from bright WR and O stars. \citep{thu05} studied
in detail the kinematics of the ionized gas in IC~10 and determined the gas velocity dispersion in
various morphological structures: shells, regions around Wolf-Rayet stars, and interstellar voids. 
Thus, the hydrogen observations also confirm the activity of star formation processes in IC~10.

\section*{OBSERVATIONAL DATA AND STELLAR PHOTOMETRY}
\indent \indent Figure 1 presents a DSS2 image of IC~10 with the labeled fields of various telescopes and detectors
that we used. To study the stellar structure of this galaxy, we used both ground-based observations 
with the 6-m BTA telescope and the 1-m Zeiss telescope at the Special Astrophysical Observatory and 
archival data from the Hubble Space Telescope (HST) with the WFPC2, ACS/WFC, and STIS cameras. Since
the images from the 1-m Zeiss telescope and several fields from the 6-m BTA telescope were used for the
most part to search for variable objects in IC~10, they are not marked in Fig. 1. General information
about the observations is given in Table 1. For several areas of the galaxy, the images with different detectors overlap: WFPC2 and ACS/WFC, ACS/WFC and 6-m BTA, STIS, and ACS/WFC. This allows us to compare the
photometric results and, if necessary, to identify objects with a variable luminosity. The WFPC2 and ACS/WFC 
images were used to study the spatial distribution of stars with various ages in IC~10 and to reveal star 
clusters, while the observations with the STIS camera and the 6-m BTA telescope (fields ST2, ST3, F5, and F7) 
were used to determine the boundary of the stellar subsystem -- the galactic thick disk.

The photometric reduction of both ground-based  and space images was performed with the
DAOPHOT II package \citep{stet94} in MIDAS. For the HST ACS/WFC images, we used the
DOLPHOT package \citep{dol00a,dol00b} for stellar photometry, which allowed us to compare
the photometric results of the two programs and to check the accuracy of our results. We subjected the
images from the ground-based BTA telescope to a  standard primary reduction and transformed instrumental
 magnitudes to Kron-Cousins magnitudes based on the equations derived during the photometry
of standard stars from the list by \citep{lan92}.

Figure 2 presents the results of our stellar photometry in the form of Hertzsprung-Russell (color-
magnitude or CM) diagrams. Since the CM diagrams for some fields are almost identical, we presented
only the most characteristic results. Whereas Fig. 2a presents the diagram for the entire ACS1 field, Fig. 2b
presents the results only for the part of the ACS2 field that we used to determine the distance. The diagrams
for the full ACS2 and ACS3 fields differ little from the diagram for the ACS1 field. There is an analogous
similarity for fields F5 and F7 of the 6-m BTA telescope and for fields ST1, ST2, and ST3 of the HST
STIS camera.

Branches of young stars (blue and red supergiants), a densely populated and wide branch of red
giants, and a branch of AGB stars are seen on all diagrams for the central regions of IC~10, especially
on the deep ACS/WFC ones. In general, if the strong reddening of all stars is disregarded, then the CM 
diagram for IC~10 does not differ in any way from the CM diagrams for irregular starburst galaxies: NGC 1569,
NGC 4214, Ho II, and others.

\section*{EXTINCTION IN IC~10}

Since, as has already been said, an accurate determination of the extinction is a key to measuring the
galaxy's parameters, let us consider the measurement of this quantity in more detail. A great variety of
methods for estimating the extinction toward IC~10 can be found in publications. As a rule, the values
obtained were used to determine the distance to the galaxy.

Since gas-dust clouds are clearly seen in the images of IC~10, it is highly likely that there is light
absorption and scattering in the galaxy itself, which must vary over the body of the galaxy quite nonuniformly.
 This means that, apart from the extinction in gas-dust clouds of our Galaxy (external component),
there must be extinction in the galaxy IC~10 itself (internal component). It is the combined action of
these components that creates the complex pattern of extinction that, as will be shown below, is observed
in IC~10.

Since the stars owe their origin to the existence of gas-dust clouds, it can be assumed that the extinction
in IC~10 will roughly correspond to the apparent distribution of young stars that have not yet
receded far from their parent gas-dust complexes. Figure 3 shows the apparent distribution of blue and
red supergiants in the central regions of IC~10. Large nonuniformities in the distribution of young stars can
be seen in the figure. This confirms the well-known fact that the stars originate in individual complexes
and points to the possible places of clumpy inhomogeneities leading to extinction fluctuations. However,
this method cannot reveal any gas-dust clouds on the periphery of IC~10 outside star-forming regions
or those in our Galaxy in the path of light propagation from IC~10.

Our study of the stellar subsystems in irregular galaxies showed that thick stellar disks with
a monotonic number density distribution of their constituent stars, red giants, exist in all galaxies
\citep{tikh05,tikh06}. Thus, the spatial distribution of red giants in IC~10 must be fairly smooth, without
large number density fluctuations, because judging by its morphology, IC~10 belongs to irregular galaxies.
This fact can be used to reveal the fluctuations of light-absorbing matter, provided that it is located
between the disk of IC~10 and the observer. However, if we construct the apparent distribution of red giants
located on a particular CM diagram (Fig. 2), then the fluctuations in the number density of stars will have
a very low contrast against the general background due to the effect of extinction in gas-dust clouds,
since the passage of light through gas-dust clouds leads to its reddening and attenuation and displaces
the star on the CM diagram but does not bring out the star far outside the red giant branch. This means
that almost all of the red giants, except the faintest ones, will remain in the list and will be involved
in constructing the apparent distribution of stars. To see the distribution of precisely those stars that
underwent an excessive reddening compared to the mean value for the entire galaxy, we should choose
the stars located at the ,,red'' edge of the red giant branch and construct their apparent distribution. In
choosing such stars on the CM diagram, we used the following parameters: $23 < I < 25$ and $(I + 4\times
(V-I) - 33) > 1$. The luminosity was constrained to remove faint stars with low photometric accuracy and
AGB stars with a spatial number density distribution differing in shape from that of red giants and the
expression in parentheses separates out the necessary part of the red giant branch. The distribution of
excessively reddened stars constructed in this way is presented in Fig. 4. As expected, an increase in
the concentration of such stars (i.e., the places of highest reddening) near the regions of young stars
can be seen, but this dependence is rather weak. A more interesting result is the appearance of zones of
strong extinction outside the concentration regions of young stars. These dark zones correspond either
to gas-dust clouds on the periphery of IC~10 or to clouds in our Galaxy (which is unlikely). In Fig. 4,
we clearly see a filamentary-clumpy pattern of lightabsorbing dark clouds even within the small (in size)
ACS/WFC field.

The deviation of the extinction from its mean value for various regions of IC~10 can be estimated using
again red giants. Constructing the luminosity function of red giants for various regions of the galaxy
and determining the beginning of the red giant branch (TRGB jump) on the diagrams, we obtain the pattern
of change in the position of the TRGB jump from region to region due to the differences in extinction. 
It should be noted that if the matter absorbing the transmitted light is located within the thick disk
of IC~10 whose stars we use, then the TRGB jump in the luminosity function will be smeared, because
there will be stars undergoing different extinctions in our sample.

Choosing small (in size) regions in the places of high and low concentration of reddened stars, i.e.,
in the zones of maximum and minimum extinction, we constructed the luminosity functions for them and
found that the difference in extinction in the I band reached $I = 1\fm3$ in them. This leads to an 
additional reddening reaching $E(V-I) = 0\fm9$. In many cases, the TRGB jump was smeared, which can be
explained by the location of the red giants and lightabsorbing matter in the same spatial volume of the
galaxy. The maximum extinction in IC~10 is observed in a small dark nebula at the southeastern edge of
the galaxy. The main-sequence stars are successively shifted there to $(V-I) = 2\fm4$, corresponding to an
extinction of $6^m$ in the V band.

These results show the entire complexity of extinction measurements, irrespective of the method used.
Therefore, it is not surprising that the distance measurements for IC~10 gave a spread in values from 0.25
to 3.0 Mpc. Summarizing the results, we can say that the extinction toward IC~10 estimated from observations 
depends noticeably on the place of its measurement and we can talk only about its local value.
Moreover, the significant extinction fluctuations over the body of the galaxy cast doubt on the results of
distance measurements by some methods, since objects located in regions with both high and low extinctions 
are involved in the measurement process. This is primarily true for the method of Cepheids, when a
small sample of stars is used, and for the method of brightest stars.

Comparison of the map of the H I distribution (\citep{wilc98}) with our map of excessively 
reddened red giants (Fig. 4) shows only their partial similarity. The filaments of excessive reddening
in Fig. 4 coincide in position with the H I filaments, but the region with highly reddened stars at the edge
of Fig. 4 does not correspond to any isolines of the H I distribution in IC~10. At the same time, we see
in the composite color image of IC~10\footnote{http//www.lowell.edu/users/massey/lgsurvey.html}
 that this region is only part of a more extended dark ring structure
 around the starburst region. The global inhomogeneities of light-absorbing matter can be seen in the
infrared DSS2 images, but no small local inhomogeneities comparable in size to those in Fig. 4 are seen
in them.

\section*{DISTANCE DETERMINATION}

The long-established fact that IC~10 belongs to the Local Group (\citep{vau65})
is beyond question, but the precise distance to this galaxy still remains a variable quantity. Table 2 gives
the distances obtained by different methods and using  different extinctions. We see from the results
presented in Table 2 that the authors often overestimate the accuracy of their measurements.
The long-used method of Cepheids could give a reliable distance to IC~10, but only if a large number
of Cepheids is used in order to average the individual extinctions. The results of distance measurements
where the extinction is calculated on the basis of one or two Cepheids (\citep{saha96,wils96})
may be considered to be nothing more than approximate estimates, bearing in mind the large extinction 
fluctuations. Even when ten Cepheids are used (\citep{sak99}, a scatter of points (up to $2^m$)
is observed on the period-luminosity diagram and, in the long run, only two or three Cepheids determine 
the distance modulus being obtained. To achieve agreement between their result and the result based
on the TRGB method, the authors have to arbitrarily introduce an additional extinction at the galactic 
center. An extreme example of inappropriateness of the method of Cepheids can be seen in the publications
by \citep{wils95} and \citep{wils96}, where the distance estimate for IC~10 changed from 240 to
820 kpc.

Using red supergiants (\citep{kar93,bor00,ovch05})
to determine the distance to IC~10 seems doubtful, because the luminosity of the brightest red supergiants
depends strongly on their metallicity and individual extinction, not to mention the calibration accuracy of 
the method itself. The as yet inadequately developed method of carbon stars used by  \citep{dem04}
also has critical remarks on the correction of extinction, the luminosity of the objects used, 
and the effect of metallicity.

Using the luminosity function of red giants and determining the cutoff point of the giant branch (the
TRGB method by \citep{lee93}) gives a reliable distance, but only under one condition: if the extinction
is known for these stars. The extinction fluctuation in the galaxy's central regions causes the position of
ITRGB to change from $21\fm8$ to $23\fm0$ from region to region. This makes the results by \citep{tikh99} and
\citep{sak99} improper, since the TRGB method was used by these authors for large areas of the
galaxy.

We attempted to circumvent these causes of in appropriateness of the TRGB method and, for this
purpose, chose several star-forming regions in IC~10 with a large number of blue supergiants. Based on the
change in the position of the reddened blue supergiant branches on the diagram relative to their normal
position, which was determined by the use of isochrones with $Z = 0.008$ and $Z = 0.02$, we calculated the 
extinctions for specific regions that we used for the subsequent work with red giants. Such a method
was applied by \citep{mas95} for the entire area of IC~10. Our distinction was that we used
images not from ground-based telescopes but from the HST. This made it possible to construct a CM diagram 
for any region of the galaxy, even dense star complexes unresolvable into stars by ground-based
telescopes. In addition, during each measurement, we used a very small region of the galaxy to reduce the
extinction fluctuations and determined the shift of the blue supergiant branch relative to its normal position
not on the basis of visual estimates but by fitting isochrones from \citep{ber94} with different
metallicities into the CM diagrams, while calculating not only the extinction but also the metallicity and age
of the star complex.

During our measurements, it emerged that not all regions of young stars could be used to determine
the distance by this method. Four of the 20 chosen regions had an unblurred TRGB boundary, 
indicating that, in most cases, the red giants and gas-dust matter are located in the same volume of the galaxy.
For each region, we constructed the luminosity function of red giants and found the cutoff point of
the red giant branch (TRGB jump) and its color index. Correcting the magnitudes for extinction and
using equations from \citep{lee93}, we found the distance moduli for the galaxy for various regions
containing clusters (Table 3) in regions containing clusters:

N20 : $E(V-I) = 1.50$, $(m-M) = 24.48$, $[Fe/H] = -1.28$, $D = 786 \pm 50$ kpc;

N36 : $E(V-I) = 1.21$, $(m-M) = 24.75$, $[Fe/H] = -1.40$, $D = 890 \pm 60$ kpc;

N48 : $E(V-I) = 1.04$, $(m-M) = 24.65$, $[Fe/H] = -1.22$, $D = 854 \pm 50$ kpc;

N54 : $E(V-I) = 1.05$, $(m-M) = 24.45$, $[Fe/H] = -0.93$, $D = 780 \pm 50$ kpc.

Since the extinction gradient undesirable for the method of distance determination is observed over
the entire apparent body of the galaxy, we also used WFPC2 images obtained away from the galactic
center (Fig. 1) to determine the distance. A densely populated branch of giants and a weak branch of
blue stars can be seen on the CM diagram for this field (Fig. 2f). The images of this region were used
by \citep{san07} to determine the distance. Applying the TRGB method, we first checked the
region for extinction fluctuations and, for our measurements, chose only part of the image at 
($0 < X < 600$ pixels), where the number density of excessively reddened stars was almost constant over
the field, indicative of a constant extinction. Since the region of measurements is far from the galactic center and
since no star-forming regions are observed nearby, while the distribution of faint blue stars is uniform
over the field, we believe that this branch of blue stars consists of old stars rather than main-sequence
ones, as was assumed by \citep{san07}. The same old blue stars are seen in large quantities in
the central regions of M31 and M32, where no star formation processes are observed. In the evolutionary 
sequence, these blue stars are low-mass core-helium-burning stars with ages of several Gyr. The
old age of IC~10 indicates that such stars can exist in this galaxy, since we found stars in the blue part
of the horizontal branch for several old globular clusters in IC~10, which can be observed only in
old star clusters. The color index of the branch of these blue stars can be found using isochrones from
\citep{ber94}. For the luminosity constraints $1.5 < M_I < 0$ and metallicity $Z = 0.008$ and $Z =
0.004$, we found the color index of such stars with ages of 7--9 Gyr to be $(V-I) = -0.33$. Taking
this value, determining the position of the reddened branch of such stars at $(V -I) = 0.68$, and using
relations from \citep{kar93}, we found the extinction in this direction: $A_I = 1.43$. For the red
giants of the region, we determined the TRGB jump in the luminosity function and the color index of the
red giant branch: $I_{TRGB} = 21.87$, $(V-I){-3.5} = 2.47$, and $(V-I)_{TRGB} = 2.8$. Using equations
from \citep{lee93}, we determined the metallicity of red giants and the distance to IC~10 in 
region S3 (Fig. 1): $E(V-I)$ = 1.01, $[Fe/H] = -1.28$, $(m-M) = 24.44$, $D = 770\pm50$ kpc.

The drawback of the result obtained is the small number of blue stars used to determine the extinction.
Therefore, we attempted to apply this approach for a field area in the ACS/WFC image closer to the galactic 
center. We chose small areas in the ACS/WFC images where there were virtually no extinction gradient 
and where no star-forming regions were observed. Only one area of field A2 located at the greatest dis-
tance from the galactic center satisfies all these conditions: $2500 < X < 3500$ pixels and $Y > 2800$ pixels.
 For this area, we found that the faint blue stars lie at $(V-I) = 0.83$. Since the normal color index
of these stars is $(V-I) = -0.33$, the reddening is $E(V-I) = 1.16$ and the extinction is $A_I = 1.64$. The
TRGB jump in the luminosity function of red giants lies at $I_{TRGB} = 22.03$, while the color of the
red giant branch is $(V-I)_{-3.5} = 2.65$ and $(V-I)_{TRGB} = 2.95$. Using equations from \citep{lee93}, 
we found for this area: $E(V-I) = 1.16$, $[Fe/H] = -1.19$, $(m-M) = 24.42$, $D = 765\pm50$ kpc.

Thus, we obtained distances for six fields of the galaxy. Since the median averaging of the results
gives a more accurate value than the simple average, applying it for our six values yields the final results:
$[Fe/H] = -1.22$, $(m-M) = 24.47$, $D = 780\pm40$ kpc. The derived distance to IC~10 indicates that the
galaxy is located at the same distance as the massive galaxies M31 and is away from it only in the plane of
the sky. Comparison of the metallicity of red giants in IC~10 with the metallicity of giants in other 
irregular or low-mass spiral galaxies shows that IC~10 is not a metal-poor galaxy but more likely has an
 enhanced metallicity.

\section*{STAR CLUSTERS}

For many years, it has been unknown which types of star clusters are present in IC~10 and whether they
exist there at all. \citep{kar93} pointed out seven candidates for star clusters in IC~10
and \citep{geor96} provided photometric data on them. Based on HST ACS/WFC images, we
checked these possible clusters and found that four of them are actually star clusters, while three probable
clusters are outside the available HST images. Based on HST WFPC2 images, \citep{hun01} pointed out
13 star clusters and associations with various ages and masses. A check of the list by \citep{hun01}
based on deep ACS/WFC images showed that the two clusters identified by this author (4--6 and 4--7)
 are single stars, while clusters 4--1 and 4--2 are most likely parts of the same extended star complex.
At the same time, the old globular cluster, N56 in our list, clearly seen in WFPC2 was not included in the
list. We found no other papers on the search for star clusters in IC~10.

Using the Large Magellanic Cloud (LMC) as an example, we see that irregular galaxies can contain
both young star clusters with various masses and old clusters (\citep{baa66}). Since the physical param-
eters of IC~10 are close to those of the LMC, we also expected to find a similar variety of star clusters
in IC~10, a galaxy with a central starburst and an extended subsystem in the form of a thick disk of old
stars (\citep{tikh99,mag03,dem04}).

Based on a visual examination of the images for IC~10, primarily fields A1 and A2 (Fig. 1), we
identified more than 50 clusters with various masses and ages. To avoid the errors in identifying poor 
clusters, we took into account the diffuse halo around the cluster, better seen in the V images, that 
is produced by faint unresolvable main-sequence stars in young clusters or by faint red giants and 
horizontal-branch stars in old clusters. Naturally, the completeness of our sample decreases when
identifying poor and old clusters; not the visual method but a computer algorithm should be used for 
their identification. The results of our search for clusters are presented in Table 3, which gives 
their coordinates, the sizes of their visually seen part, and morphology. The true sizes of the star
complexes can exceed significantly those given in Table 3, because their periphery is lost among the 
numerous background stars.

The images that we used cover almost the entire central part of IC~10 with star-forming regions near
which young clusters are located, but several more young clusters can be found in the western part of
the galaxy that is not covered by the HST images. As regards the old globular clusters, the discovery of 
several clusters is also possible here, because much of the galactic periphery remains unstudied. The 
example of the WFPC2 image (field S2 in Fig. 1) in which an old globular cluster was found on the galactic
periphery is revealing.

All of the star clusters found can be separated into young, old, and intermediate-age ones. In turn,
the young clusters can be separated into star complexes, bright globular clusters, and open or compact
poor clusters. The old clusters can be separated into bright globular and fain clusters, while
the intermediate-age clusters have a low brightness and contain faint main-sequence stars and faint red
supergiants. Within the HST ACS/WFC images, we found: 8 young complexes, 10 young open clusters,
12 young compact clusters, 5 young globular clusters, 5 intermediate-age clusters, 5 old open clusters,
 and 12 old globular clusters. In this list, one old globular cluster (N1) was found in the WFPC2
image (field S2) and one old globular cluster (N49) was taken from \citep{kar93}.
Thus, IC~10 has star clusters of the same types as those in the LMC, which confirms a morphological
similarity of these galaxies, except the slightly lower mass of IC~10.

When separating the clusters by their age into young and old ones, we used stellar photometry and
CM diagrams for the clusters found. Some of the clusters have a high star density and the stellar
photometry of such clusters is possible only on their periphery. In these cases, the CM diagram for
a cluster can contain a significant number of background stars from the disk of IC~10 that are projected 
onto the cluster region. The resulting mixture of stars with various ages creates an uncertainty in the
separation of the clusters into young and old ones. To make the separation procedure more objective, we selected
a strip in the galactic image with the center in the cluster under study and plotted the number density
distribution of old or young stars along this strip. If the cluster is young, then the number density of young
stars will peak on the plot at the point corresponding to the position of the cluster center. A similar picture
will be observed when the distribution of red giants or horizontal-branch stars is constructed for an old
cluster. The ACS/WFC images of several young and old clusters found are shown in Fig. 5. One of the old
globular clusters (N10) is very elongated in shape. This can be explained either by an optical superposition
of two clusters, which is unlikely, or by the result of a gravitational interaction between a cluster
and the galaxy. Note that similar elongated globular clusters are present in the LMC, thereby confirming a
morphological similarity of the two galaxies.

Having the tables of stellar photometry for IC~10 at our disposal, we constructed CM diagrams for
several young complexes and compared them with isochrones from \citep{ber94}. Since the
sizes of all complexes do not exceed several arcsec, the extinction may be assumed to be approximately
the same for all stars of the same complex. This is confirmed by the small width of the blue supergiant
branch in each complex. On the derived diagrams, we see that the color index of the blue supergiant
branch changes from complex to complex, which directly indicates that the extinction is different for
different regions of the galaxy. In their measurements, \citep{leq79} and \citep{gar90} obtained a
low metallicity ($Z = 0.3Z_{\sun}$) for H II regions in IC~10. Therefore, it can be assumed that the isochrones with
$Z = 0.008$ will describe well our CM diagrams. However, comparison showed (Fig. 6) that the isochrones
with $Z = 0.02$, i.e., with a solar metallicity, provide the best fit. Since this comparison was made for
several star complexes containing red supergiants, it can be argued that IC~10 is not a metal-poor
galaxy, as has been assumed up until now. Having identified bright stars with large color indices (among
which there can also be red dwarfs of our Galaxy), we constructed their apparent distribution over the body
of IC~10. The concentrations of such stars coincident with star-forming regions or compact clusters can
be seen in the derived distribution. This proves that these stars belong to bright red supergiants but not
to foreground dwarfs of our Galaxy. Thus, it should be recognized that the brightest supergiants with ages
of 15-30 Myr in many young clusters of IC~10 have a solar metallicity.

The high metallicity of stars in IC~10 that we found may remove the contradiction between the excessively 
large ratio of the numbers of carbon and nitrogen (WC/WN) Wolf-Rayet stars and the low
metallicity of the galaxy assumed so far. Because of the large number of clusters found, their detailed
study is a matter of a separate paper that we are planning to present in the immediate future.

\section*{THE SIZES OF THE GALACTIC THICK DISK AND HALO}

Whereas IC~10 has an irregular shape in the visual spectral range, a disk with a regular shape that
consists mostly of red stars and has an axial ratio a/b = 1.38 is seen on the 2MASS infrared ($K_s$-band)
images. Clearly, we see only the central part of the disk, where AGB stars with ages from 100 to
800 Myr make the largest contribution to its emission. To see the distribution of old RGB stars that
constitute the thick disk of IC~10, it is necessary to identify these stars on the CM diagram and to 
construct their apparent distribution over the body of the galaxy. Unfortunately, the insufficient angular sizes of
the available images and the extinction fluctuations over the body of the galaxy lead to a distortion of
the actual distribution of stars, but even the available data indicate that the stellar subsystems of IC~10
are similar in structure to those of irregular galaxies. Since the thick-disk sizes for most irregular galax-
ies determine the maximum sizes of the galaxies, it seems interesting to identify the thick disk in IC~10.
The existence of this stellar subsystem can be seen from the exponential decrease in the number density
of red giants toward the galactic edge in all of the images we used. The size of the thick disk in IC~10
was first determined by \citep{tikh99} based on images from the 6-m BTA telescope. The derived disk
diameter of 18 was minimal, because the thick-disk boundary was not reached. Subsequently, additional
images from the 6-m BTA telescope were used and the thick-disk boundary along the galaxy's minor axis
was found to pass at a galactocentric distance of 10.5 and a more extended halo stretches further out
(\citep{tikh02,droz03}). A jump in the number density of stars and a change in the den-
sity gradient are observed (Fig. 7) at the thick-disk boundary (fields F5 and F7 in Fig. 1), corresponding
to the passage from the thick disk to the halo. If the outer disk has the same axial ratio as the inner one,
then the thick disk is $29\arcmin\times21\arcmin$ in size. The thick disk that we identified is indicated 
in Fig. 1 in the form of
an ellipse. The number density distributions of stars in fields ST2 and ST3 give an additional confirmation of
the validity of the thick-disk boundary. In the former case, a simple decrease in the number density of stars
is observed; in the latter case, a change in the density gradient is seen, although these data are not very
reliable due to the small number of stars available in these fields.

The halo whose far boundary has not yet been determined is even larger in size than the thick disk.
Based on the distribution of red giants, \citep{dem04} found the maximum sizes of IC~10 to
be $32\arcmin$ . In their figure, the distribution has the shape of a horseshoe, which is the result of an inhomogeneous
extinction in IC~10 and a low photometric limit of the images used. It is beyond doubt that the halo of
IC~10 has larger sizes than those of the galaxy given by \citep{dem04}.

\citep{huch79} and \citep{coh79} found the hydrogen disk of IC~10 to be more than one degree
in size. Does the stellar halo extend to the same distance? Solving this question can also give an answer 
to the question about the origin of the hydrogen cloud around the galaxy. Does the cloud arise (at least
partially) from the mass loss by stars or it has an external origin, forming when a remote gas falls to the
galaxy? For many of the galaxies we investigated, the sizes of the hydrogen cloud have never exceeded those
of the stellar disks or halo, including, for example, those for such an irregular galaxy as NGC 2915 with
apparent sizes of $1.\arcmin0\times1.\arcmin9$, while the hydrogen disk is at least $20\arcmin$ with a
 distinct spiral structure. When the
stellar subsystems of IC~10 are studied, its fairly close neighbor, the giant spiral galaxy M31, should be kept
in mind. The gravitational interaction between the two galaxies can change significantly the morphology
of IC~10 even if this interaction took place in the distant past.

\section*{RESULTS AND CONCLUSIONS}
\indent \indent For several regions of IC~10, we performed deep stellar photometry based on which we constructed
Hertzsprung-Russel diagrams for stars of both central regions and periphery of the galaxy. The derived
CM diagrams for IC~10 do not differ in any way from those for irregular starburst galaxies. Distance
moduli with a median value of $(m-M) = 24.47$, $D = 780 \pm 40$ kpc, were obtained for six galactic fields by
the TRGB method. Our study of the distribution of red giants revealed extinction inhomogeneities reaching
 $\delta I = 1\fm3$ or $E(V-I) = 0\fm9$ within the investigated galactic fields. Such inhomogeneities create
diffculties in determining the distance to the galaxy and measuring its physical parameters. A visual ex-
amination of the images and our stellar photometry revealed 57 star clusters in the galaxy with various
ages and masses. Comparison of the CM diagrams for clusters with theoretical isochrones allowed us to
determine the ages and metallicities of cluster stars. The metallicities of young supergiants in IC~10 were
found to be nearly solar. This means that IC~10 has an enhanced metallicity, which probably also explains
the high ratio of the numbers of carbon and nitrogen Wolf-Rayet stars (WC/WN). We determined the size
of the thick disk in IC~10 along its minor axis by measuring the gradient in the number density of red
giants: $B = 10.\arcmin5$. The halo, which also consists of old stars, extends to an even greater distance. It may
well be that the size of the halo will coincide to that of the hydrogen cloud around IC~10, i.e., will have a
diameter of at least one degree.

Our results close some gaps in the investigation of IC~10 and provide a basis for further studies. In
particular, the star clusters of IC~10 should be studied in more detail to compare their parameters with those
of clusters in other irregular galaxies. The questions about the sizes of the stellar halo in IC~10 and the
structure of its outer periphery still also remain unanswered, since the possible interaction of IC~10 with
M31 could change the morphology of the thick disk and halo in IC~10, particularly their outer parts. The
high metallicity of young stars in IC~10 that we obtained requires accurate spectroscopic observations
of young supergiants to confirm these results, which may well be performed in the immediate future.

\begin{table}[t]
\begin{center}
\footnotesize
\renewcommand{\tabcolsep}{13pt}
\caption{Observational data.}
\vspace{0.2cm}
 \begin{tabular}{lllccc} \hline
 \multicolumn{1}{c}{Date}&
 \multicolumn{1}{c}{Telescope}&
\multicolumn{1}{c}{Field}&
\multicolumn{1}{c}{Filter}&
\multicolumn{1}{c}{Exposure}&
 \multicolumn{1}{c}{Proposal ID}\\ \hline
  June 1997   &  HST WFPC2   &  S1 & F814W     &   1400 $\times$ 10  & 6406\\
  July 1998   &  6--m BTA     &  F5 & I         &    600               &          \\
  July 1998   &  6--m BTA     &  F5 & R         &    600               &          \\
  June 1999   &  HST WFPC2   &  S1 & F555W     &   1400 $\times$ 10   & 6406\\
  June 1999   &  HST STIS    &  ST1& (V)$^*$   &  1200 $\times$  2    & 7912\\
  June 1999   &  HST STIS    &  ST1& (I)$^*$   &  1200 $\times$  2    & 7912\\
  Jan. 1999   &  6-m BTA     &  F1 & I         &    600               &          \\
  Jan. 1999   &  6-m BTA     &  F1 & R         &    600               &          \\
  Oct. 2000   &  6-m BTA     &  F7 & I         &  300 $\times$3       & \\
  Oct. 2000   &  6-m BTA     &  F7 & R         &  300$\times$ 3       & \\
  Oct. 2002   &  HST STIS    &  ST2& (V)$^*$   &  480 $\times$  2     & 9633\\
  Oct. 2002   &  HST STIS    &  ST2& (I)$^*$   & $\Sigma$ = 800       &  9633\\
  Oct. 2002   &  HST STIS    &  ST3& (V)$^*$   &  480 $\times$  2     & 9633\\
  Oct. 2002   &  HST STIS    &  ST3& (I)$^*$   & $ \Sigma$ = 800      &  9633\\
  Oct. 2002   &  HST WFPC2   &  S2 & F606W     &    500 $\times$ 2    &   9676\\
  Oct. 2002   &  HST ACS/WFC &  A1 & F606W     &    2160              & 9683\\
  Oct. 2002   &  HST ACS/WFC &  A1 & F814W     &    2160              & 9683\\
  Oct. 2002   &  HST ACS/WFC &  A2 & F606W     &    1080$\times$ 2    & 9683\\
  Oct. 2002   &  HST ACS/WFC &  A2 & F814W     &    1080$\times$ 2    & 9683\\
  Jan. 2005   &  HST WFPC2   &  S3 & F555W     &    500 $\times$ 12   &  10242\\
  Jan. 2005   &  HST WFPC2   &  S3 & F814W     &   500 $\times$ 12    &  10242\\
  Jan. 2005   &  HST ACS/WFC & A3  &  F555W    &    2480 $\times$ 8   & 10242\\
  Jan. 2005   &  HST ACS/WFC & A3  &  F814W    &    2380 $\times$ 4   &10242\\
\hline
\multicolumn{6}{l}
{Note. (V )$^*$ and (I)$^*$ are broadband filters that we reduced to the international V and I bands.}\\
\end{tabular}
\end{center}
\end{table}

\begin{table}
\begin{center}
\footnotesize
\renewcommand{\tabcolsep}{6pt}
\caption{Results of distance measurements for IC 10}
\vspace{0.2cm}
 \begin{tabular}{|clcll|} \hline
\multicolumn{1}{|c}{$D$, Mpc}&
\multicolumn{1}{c}{$\sigma$}&
\multicolumn{1}{c}{$E(B-V)$}&
\multicolumn{1}{c}{Method}&
\multicolumn{1}{c|}{Authors}\\ \hline
  1.5  &  $-$   & $-$    &  H II rings                  & Roberts (1962)\\
  1.25 &  $-$   & 0.87   &  H II rings                  & de Vaucouleurs \& Ables (1965)\\
  3.0  &  0.3   & 0.78   &  H II rings                  & Sandage \& Tammann (1974)\\
  2.0  &  0.4   & 0.40   &  H II rings                  & de Vaucouleurs (1978)\\
  1.8  &  0.5   & 0.78   &  Planetary nebulae           & Jacoby \& Lesser (1981)\\
  2.0  &  $-$   & 0.40   &  Tully-Fisher                & Bottinelli et al. (1984)\\
  1.04 & 0.09   & 0.87   &  Brightest stars             & Karachentsev \& Tikhonov (1993)\\
  0.24 & $-$    & 1.55   &  Cepheids                    & Wilson (1995)\\
 0.95   & 0.09  & 0.80   &  Wolf-Rayet stars            & Massey \& Armandroff (1995)\\
 0.82   & 0.08  & 0.80   &  Cepheids                    & Wilson et al. (1996)\\
 0.83   & 0.11  & 0.94   &  Cepheids                    & Saha et al. (1996)\\
 0.58   & 0.05  & 0.98   &  Tip of the red giant branch & Tikhonov (1999)\\
 0.66   & 0.07  & 1.16   &  Cepheids                     & Sakai et al. (1999)\\
 0.50   & 0.05  & 0.85   &  Tip of the red giant branch & Sakai et al. (1999)\\
 0.59   & 0.05  & 1.05   &  Brightest stars             & Borissova et al. (2000)\\
 0.74   & 0.04  & 0.79   &  Carbon stars                & Demers et al. (2004)\\
 0.72   & 0.11  & 1.25   &  Brightest stars             & Ovcharov \& Nedialkov (2005)\\
 0.78   & 0.03  & 0.95   &  Tip of the red giant branch & Vacca et al. (2007)\\
 0.89   & 0.04  & 0.60   &  Tip of the red giant branch & Sanna et al. (2007)\\
 0.78   & 0.05  & 0.80   &  Tip of the red giant branch & This paper (2009)\\
\hline
\end{tabular}
\end{center}
\end{table}

\hoffset=-20mm
\hoffset=-10mm
\topmargin=-1cm
\renewcommand{\baselinestretch}{1}
\begin{table}
\begin{center}
\footnotesize
\renewcommand{\tabcolsep}{6pt}

\caption{Star clusters in IC 10}
\vspace{.1cm}
 \begin{tabular}{|cllcl|} \hline
\multicolumn{1}{|c}{No.}&
\multicolumn{1}{c}{$RA$}&
\multicolumn{1}{c}{$DEC$}&
\multicolumn{1}{c}{Size, arcsec}&
\multicolumn{1}{c|}{Cluster morphology}\\ \hline

01 &  {\bf00 19 53.98} & {\bf+59 13 25.7}  &  3.0 & Old globular, bright\\
02 &  {\bf00 19 57.84} & {\bf+59 19 53.3}  &  1.5 & Intermediate age, open\\
03 &  {\bf00 19 59.25} & {\bf+59 19 35.7}  &  0.8 & Young, compact, bright\\
04 &  {\bf00 20 00.16} & {\bf+59 19 58.6}  &2.8 & Old globular, faint\\
05 &  {\bf00 20 02.05} & {\bf+59 19 45.5}  &  1.5 & Young globular, poor\\
06 &  {\bf00 20 02.05} & {\bf+59 20 04.7}  & 0.6 & Young, compact, bright\\
07 &  {\bf00 20 02.22} & {\bf+59 19 07.7}  &  1.5 & Old globular, faint\\
08 &  {\bf00 20 03.19} & {\bf+59 18 50.5}  &   1.9 & Old open, faint\\
09 &  {\bf00 20 04.34} & {\bf+59 18 34.9}  &  2.0 & Old open, poor\\
10 &  {\bf00 20 05.68} & {\bf+59 18 26.5}  &  3.2 & Old globular, very elongated\\
11 &  {\bf00 20 06.57} & {\bf+59 19 22.5}  &  1.7 & Young open, poor\\
12 &  {\bf00 20 06.68} & {\bf+59 19 09.0}  &   1.2 & Old open, poor\\
13 &  {\bf00 20 07.53} & {\bf+59 19 16.3}  &  1.2 & Old globular, faint\\
14 &  {\bf00 20 07.58} & {\bf+59 19 26.7}  &  1.4 & Young, compact, bright\\
15 &  {\bf00 20 09.64} & {\bf+59 17 19.2}  &  2.3 & Young globular, bright\\
16 &  {\bf00 20 10.45} & {\bf+59 18 22.2}  &  0.8 & Young, compact, poor\\
17 &  {\bf00 20 10.47} & {\bf+59 21 07.3}  &   1.5 & Old, open, poor\\
18 &  {\bf00 20 11.50} & {\bf+59 18 50.9}  &  1.6 & Young, open, bright\\
19 &  {\bf00 20 12.37} & {\bf+59 19 16.8}  & 1.6 & Young globular, bright\\
20 &  {\bf00 20 12.40} & {\bf+59 17 27.7}  &  2.6 & Young globular, bright\\
21 &  {\bf00 20 13.92} & {\bf+59 21 14.7}  &  1.0 & Old, compact, poor\\
22 &  {\bf00 20 15.42} & {\bf+59 19 49.8}  &  0.6 & Intermediate age, compact\\
23 &  {\bf00 20 17.23} & {\bf+59 17 01.6}  & 3.4 & Old globular, bright\\
24 &  {\bf00 20 17.24} & {\bf+59 17 45.3}  &  2.8 & Young globular, bright\\
25 &  {\bf00 20 17.37} & {\bf+59 16 56.1}  &  2.7 & Young open, poor\\
26 &  {\bf00 20 17.71} & {\bf+59 19 17.5}  &  1.3 & Young open, poor\\
27 &  {\bf00 20 17.86} & {\bf+59 17 46.3}  &  0.9 & Young, compact, bright\\
28 &  {\bf00 20 17.89} & {\bf+59 17 01.9}  & 4.1 & Young, open, bright\\
29 &  {\bf00 20 18.03} & {\bf+59 19 50.5}  &  3.7 & Old globular, bright\\
30 &  {\bf00 20 18.45} & {\bf+59 17 58.4}  &  1.0 & Young, compact, bright\\
31 &  {\bf00 20 18.42} & {\bf+59 18 23.4}  &  1.4 & Old globular, faint\\
32 &  {\bf00 20 18.63} & {\bf+59 18 55.3}  &  5.5 & Young complex with bright stars\\
33 &  {\bf00 20 18.94} & {\bf+59 18 08.9}  &  1.0 & Young, compact, bright\\
34 &  {\bf00 20 19.19} & {\bf+59 17 30.3}  &  1.6 & Old globular, faint\\
35 &  {\bf00 20 20.07} & {\bf+59 18 20.9}  &  1.4 & Young, open, poor\\
36 &  {\bf00 20 20.37} & {\bf+59 18 37.7}  &  1.4 & Core of wide young complex\\
37 &  {\bf00 20 20.96} & {\bf+59 17 13.1}  &  2.0 & Young open, poor\\
38 &  {\bf00 20 21.02} & {\bf+59 18 59.5}  & 1.1 & Intermediate age, compact, poor\\
39 &  {\bf00 20 21.71} & {\bf+59 18 22.4}  &  1.0 & Young, compact, poor\\
40 &  {\bf00 20 21.79} & {\bf+59 17 41.0}  &  0.6 & Young, very compact, bright\\
41 &  {\bf00 20 22.43} & {\bf+59 17 15.5}  &  9.6 & Young complex with bright stars\\
42 &  {\bf00 20 23.09} & {\bf+59 16 52.9}  & 0.5 & Intermediate age, compact, poor\\
43 &  {\bf00 20 23.74} & {\bf+59 17 33.1}  &  1.2 & Young open, poor\\
44 &  {\bf00 20 24.21} & {\bf+59 19 10.1}  &  2.8 & Intermediate age, open, rich\\
45 &  {\bf00 20 24.51} & {\bf+59 18 18.1}  & 2.1 & Young open, poor\\
46 &  {\bf00 20 24.69} & {\bf+59 18 11.8}  &  1.0 & Young, compact, bright\\
47 &  {\bf00 20 25.04} & {\bf+59 17 39.2}  & 2.8 & Young, open, rich\\
48 &  {\bf00 20 27.36} & {\bf+59 21 14.6}  & 2.5 & Old globular, bright\\
50 &  {\bf00 20 26.76} & {\bf+59 17 02.5}  & 1.0 & Young, compact, poor\\
51 &  {\bf00 20 26.67} & {\bf+59 19 48.0}  &   1.3 & Old globular, faint\\
52 &  {\bf00 20 27.48} & {\bf+59 17 23.4}  &  2.6 & Young complex with bright stars\\
53 &  {\bf00 20 27.48} & {\bf+59 17 07.7}  &  0.4 & Young, very compact, poor\\
54 &  {\bf00 20 27.80} & {\bf+59 17 38.4}  &  3.8 & Young complex with bright stars\\
55 &  {\bf00 20 29.15} & {\bf+59 16 57.8}  &  7.2 & Young complex with bright stars\\
56 &  {\bf00 20 29.57} & {\bf+59 18 07.5}  & 1.2 & Old globular, faint\\
57 &  {\bf00 20 32.18} & {\bf+59 17 12.1}  & 1.8 & Core of wide young complex\\
\hline
\end{tabular}
\end{center}
\end{table}

\begin{figure}[t]
\centerline{\includegraphics[angle=0, width=16cm, bb=37 17 573 620,clip]{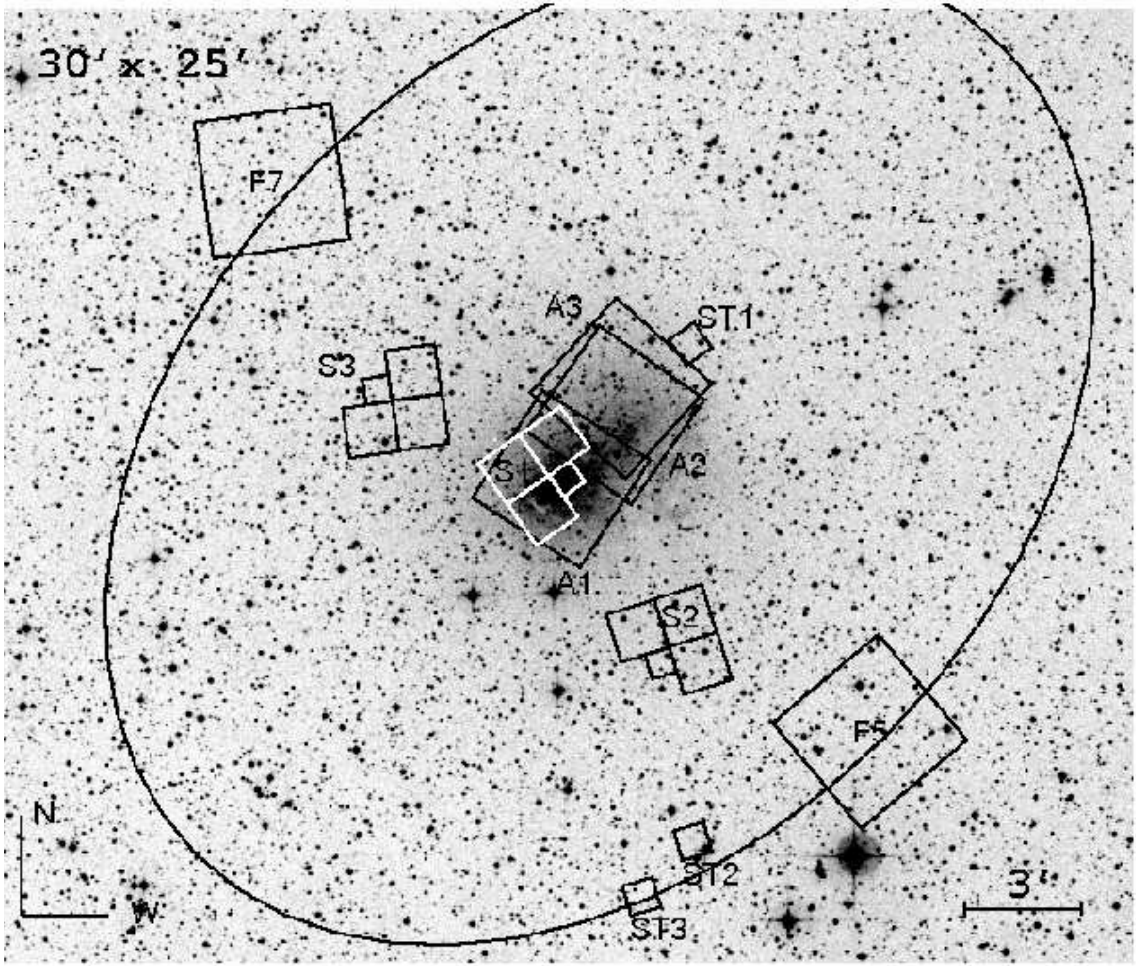}}
\caption{DSS2 image of IC 10. The fields that we used in studying the galaxy's stellar composition are labeled: F5 and F7 for
the 6-m BTA telescope; ST1, ST2, ST3 for the STIS camera; S1, S2, S3 for the WFPC2 camera; A1, A2, A3 for the HST
ACS/WFC camera. The ellipse marks the boundary between the thick disk and the more extended halo.}
\end{figure}
\begin{figure}[t]
\centerline{\includegraphics[angle=0, width=17cm, bb=25 22 607 522,clip]{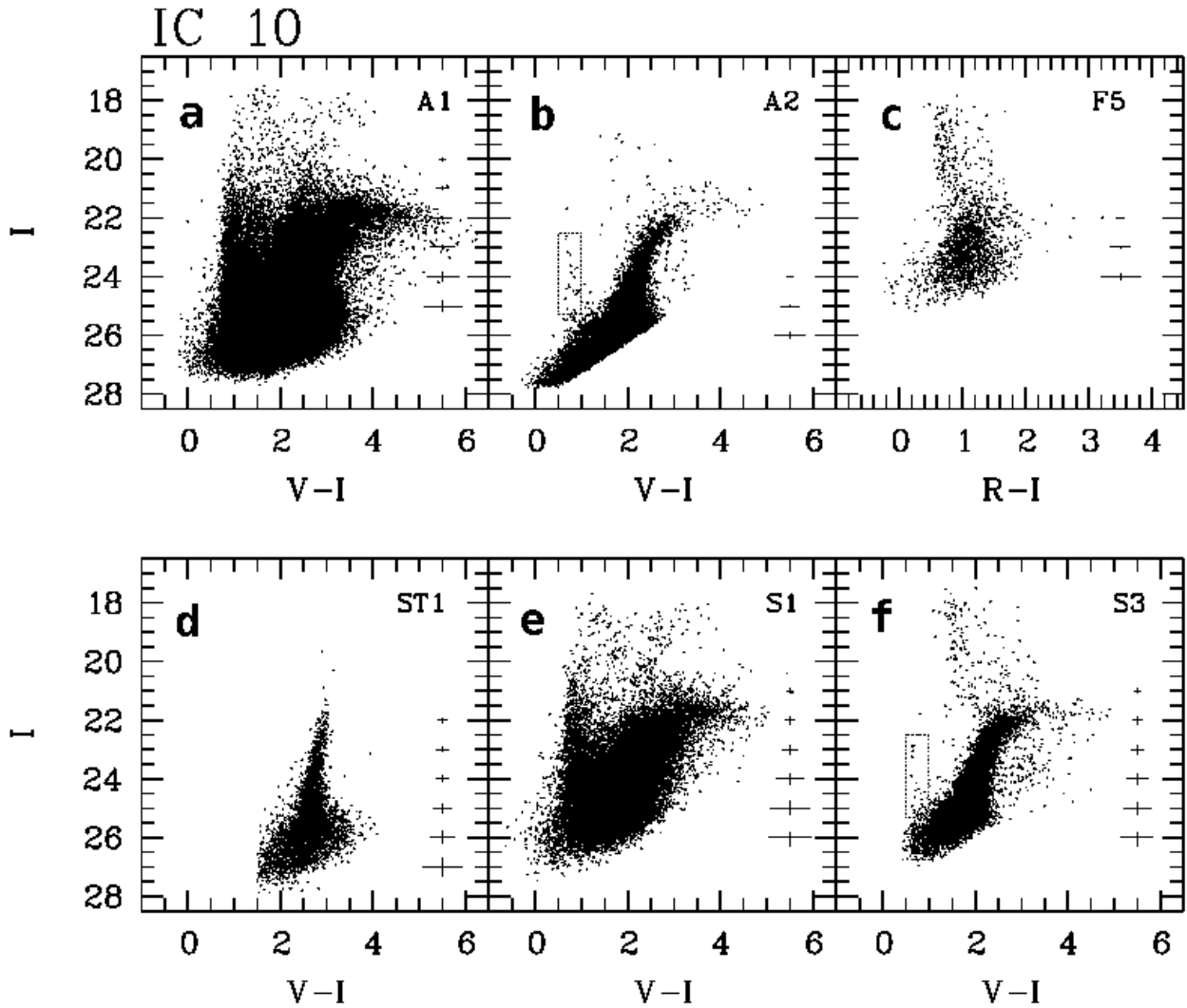}}%
\caption{Hertzsprung-Russell diagrams for various regions of IC 10: (a) and (e) star-forming regions (ACS1 and WFPC2); (b)
a field far from the galactic center (ACS2); (c) thick-disk periphery (F5); (d) and (f) fields in the thick disk (ST1 and WFPC2).
On the CM diagrams for the ACS2 and WFPC2 fields, the branches of old blue stars of these fields are marked.}
\end{figure}
\begin{figure}[t]
\centerline{\includegraphics[angle=0, width=8cm, bb=118 131 380 450,clip]{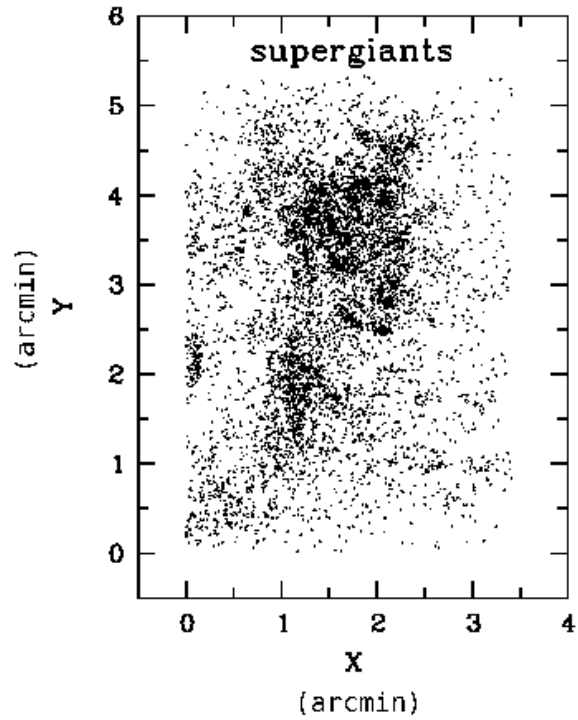}}
\caption{Apparent distribution of young supergiants in the galaxy's central regions (fields A1 and A2). The concentration zones
of young stars point to the places of a possible excessive reddening of stars due when light passes through gas-dust clouds
of IC 10.}
\end{figure}
\begin{figure}[t]
\centerline{\includegraphics[angle=0, width=8cm, bb=118 131 380 450,clip]{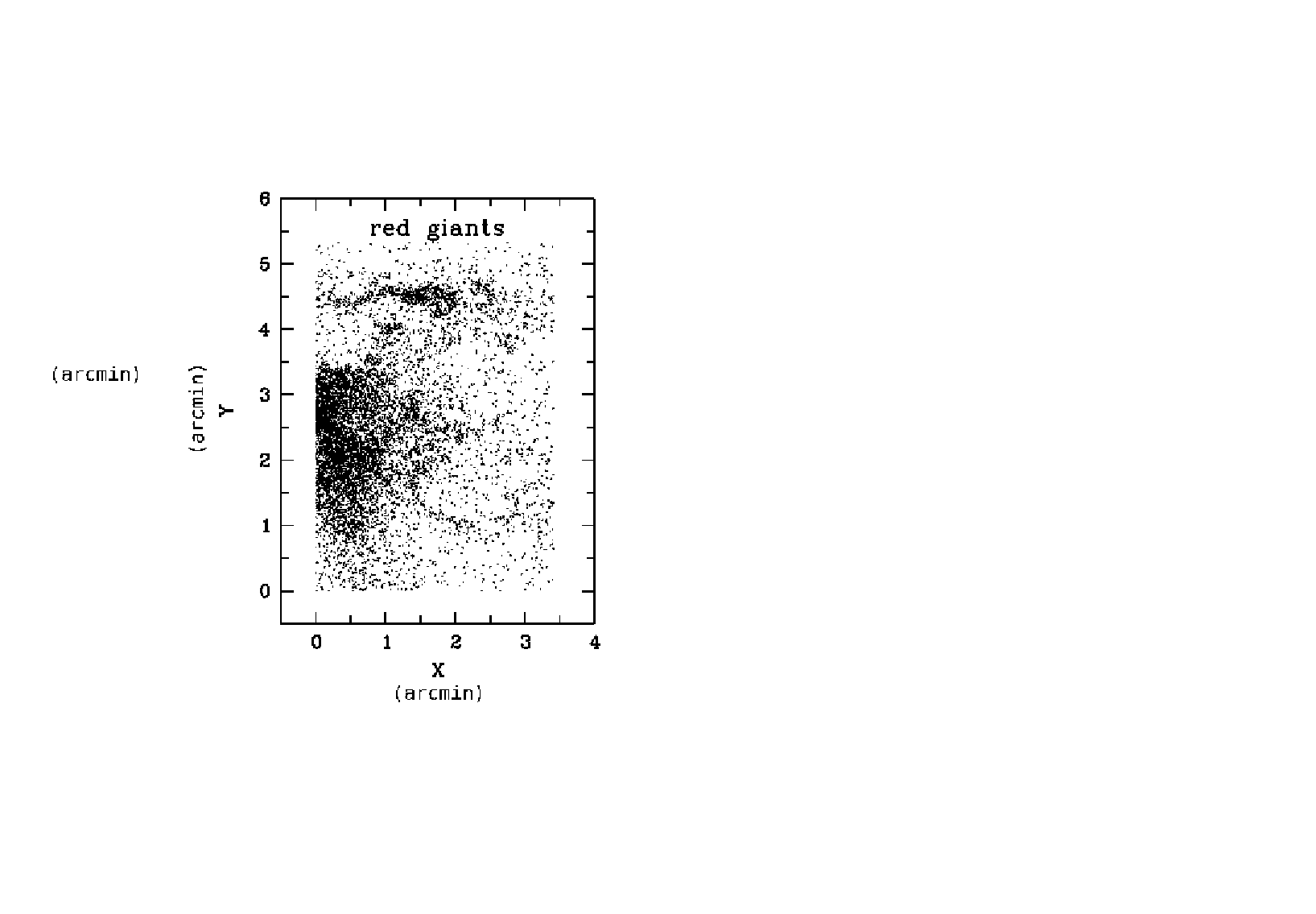}}
\caption{Apparent distribution of excessively reddened red giants. The field boundaries correspond to those in Fig. 3. The narrow
filaments with a concentration of reddened stars correspond to the positions of H I filaments. The wide region of concentration
of reddened stars seen in the left part of the figure does not correspond to any significant H I clouds, but corresponds to the dark
regions seen on the composite color image of IC 10.}
\end{figure}
\begin{figure}[t]
\centerline{\includegraphics[angle=0, width=12cm, bb=83 181 500 818,clip]{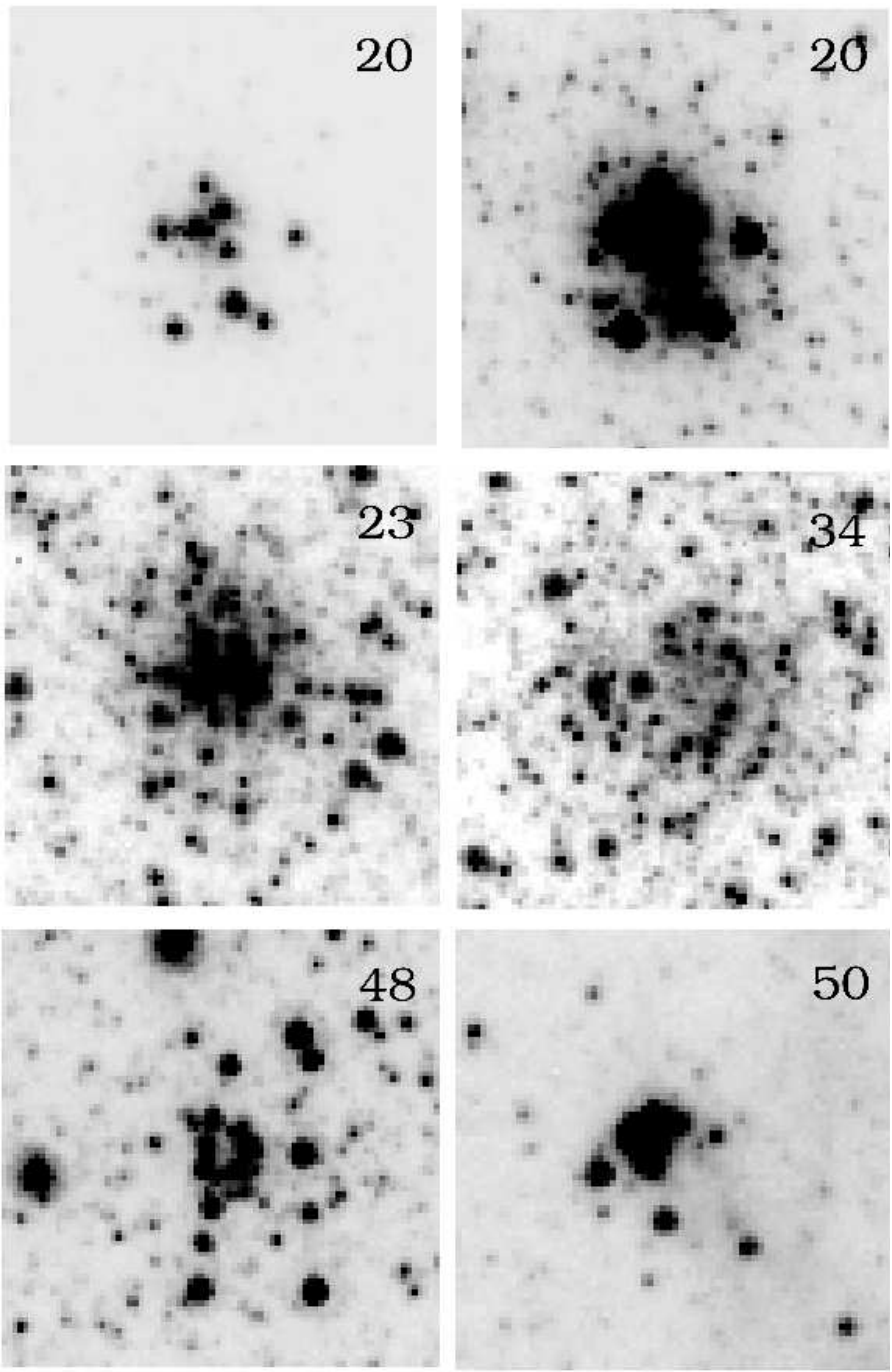}}
\caption{ACS/WFC images of several clusters in IC 10: two images of N20 with different reproduction conditions -- a young
globular cluster with red supergiants, N23 -- an old globular cluster, N34 -- a young open cluster, N48 -- the core of a wide
star complex, N50 -- a compact young cluster.}
\end{figure}
\begin{figure}[t]
\centerline{\includegraphics[angle=0, width=15cm, bb=70 165 550 825,clip]{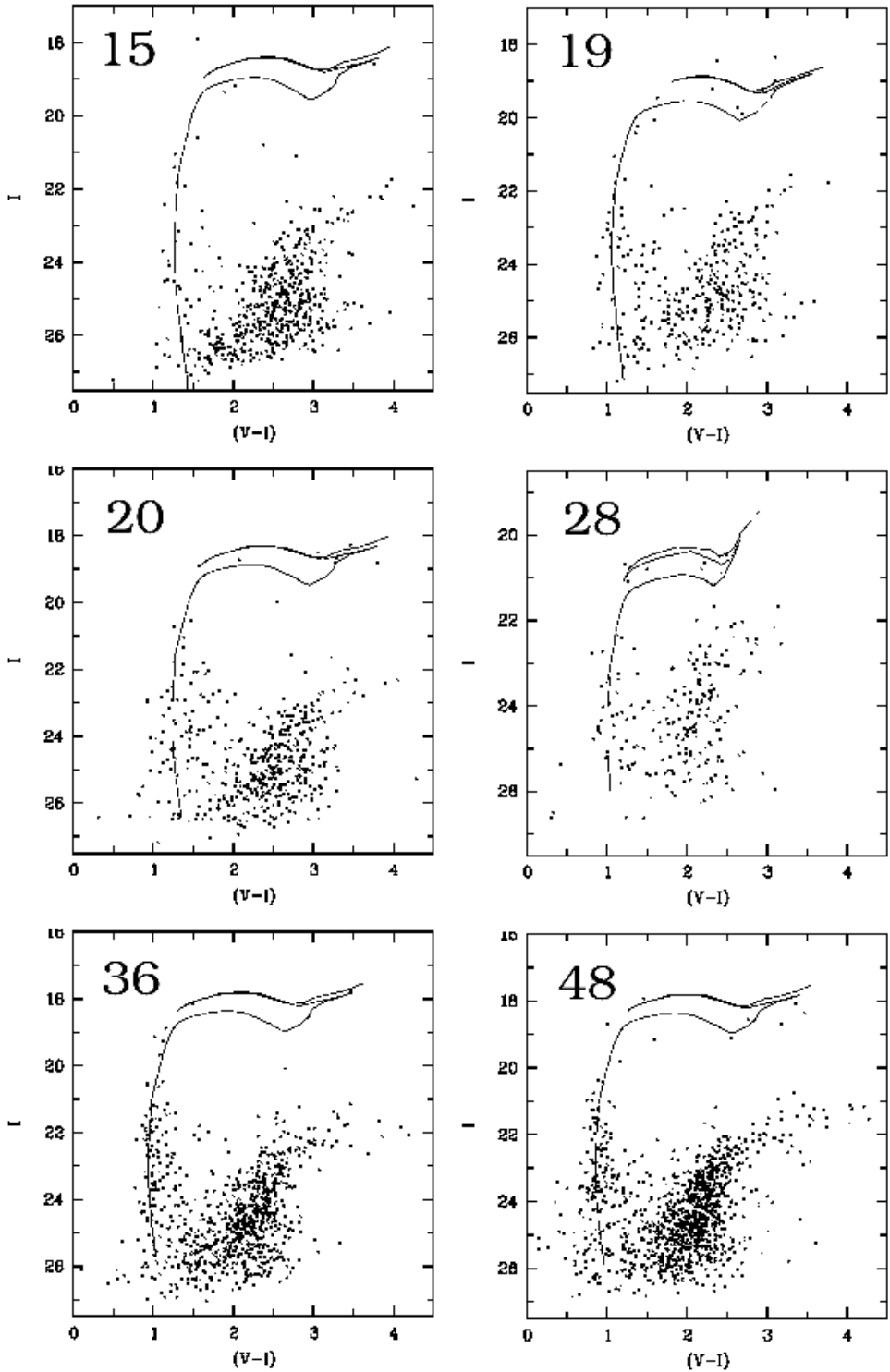}}
\caption{CM diagrams for the clusters N15, N19, N20, N28, N36, and N48 with fitted t = 13 Myr isochrones and metallicity
Z = 0.02, except N28, where the isochrone with metallicity Z = 0.008 was fitted. We see that bright red supergiants with a
high metallicity equal to the solar one are present in most of the presented clusters. On the diagrams, we see that the blue
supergiant branches have different color indices, indicating that the extinction varies from cluster to cluster.}
\end{figure}
\begin{figure}[t]
\centerline{\includegraphics[angle=0, width=15cm, bb=70 520 550 780,clip]{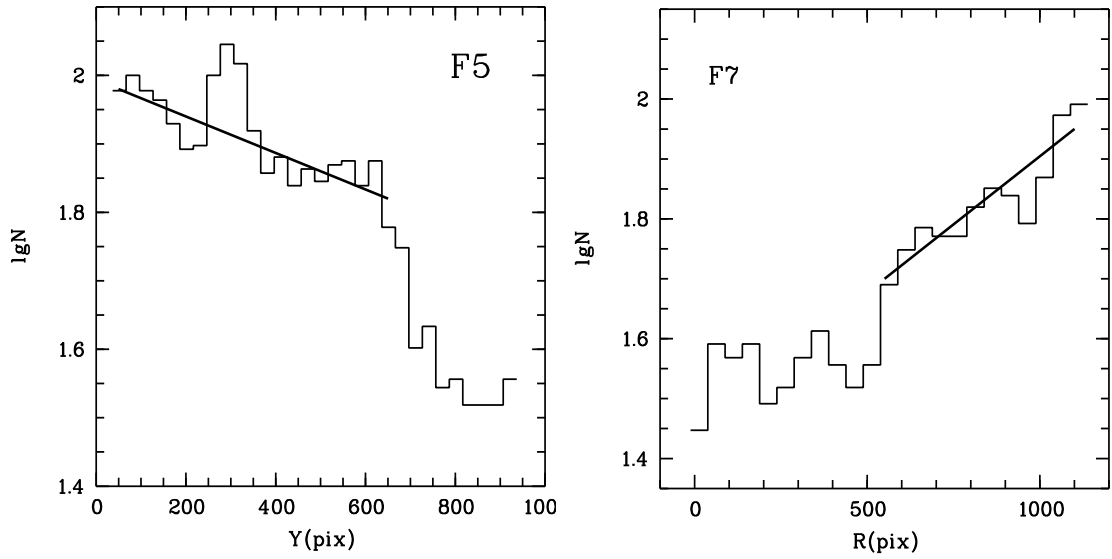}}
\caption{Change of the gradient in the number density of red giants at the boundary of the thick disk and halo in fields F5 and F7.
Outside the thick disk at B = 10.5, the gradient in the number density of red giants decreases sharply, but the halo still extends
to a considerable distance from the galactic center.}
\end{figure}
\end{document}